# Audio-to-Image Encoding for Improved Voice Characteristic Detection Using Deep Convolutional Neural Networks

*Youness Atif*

ABSTRACT

This paper introduces a novel audio-to-image encoding framework that integrates multiple dimensions of voice characteristics into a single RGB image for speaker recognition. In this method, the green channel encodes raw audio data, the red channel embeds statistical descriptors of the voice signal (including key metrics such as median and mean values for fundamental frequency, spectral centroid, bandwidth, rolloff, zero-crossing rate, MFCCs, RMS energy, spectral flatness, spectral contrast, chroma, and harmonic-to-noise ratio), and the blue channel comprises subframes representing these features in a spatially organized format. A deep convolutional neural network trained on these composite images achieves 98% accuracy in speaker classification across 2 speakers, suggesting that this integrated multi-channel representation can provide a more discriminative input for voice recognition tasks.

## 1   Introduction

Recent breakthroughs in neuroscience have revealed that the brain's sensory cortices exhibit remarkable plasticity. For example, Sur [1] demonstrated that when visual inputs are rerouted to the auditory cortex, this region—traditionally dedicated to processing sound—can adapt to interpret visual stimuli, effectively enabling animals to "see" with their ears and "hear" with their eyes. Motivated by these insights, this work proposes a novel audio-to-image encoding framework for voice characteristic detection using deep convolutional neural networks. In this approach, the green channel is dedicated to storing the raw audio data, the red channel encodes key statistical descriptors of the voice signal—such as pitch (fundamental frequency), spectral centroid (brightness), spectral bandwidth (energy spread), spectral rolloff (energy distribution), zero-crossing rate (noisiness), MFCCs (spectral envelope), RMS energy (loudness), harmonic-to-noise ratio (periodicity vs. noise), spectral flatness (tonality), spectral contrast (peak–valley differences), and chroma features (harmonic structure)—and the blue channel is subdivided into subframes that visually represent these metrics along with their corresponding medians. Data from 2 speakers (Aria and Rachid) was collected by randomly sampling 548 phrases per speaker, resulting in 1096 composite images. A CNN trained [2] on these images—enhanced through data augmentation techniques such as zooming, flipping and rotating—achieved a speaker classification accuracy of 98%. This integrated, multi-channel encoding strategy not only enables precise audio reconstruction but also provides a richer, more discriminative input for voice recognition tasks, highlighting its potential for applications in multimodal signal processing.

## 2   From Sound Waves to Pixels

Sound, in its natural state, is a continuous-time analog signal represented by a function $x(t)$. To digitize this signal, two key processes are applied: sampling and quantization. According to the Nyquist–Shannon sampling theorem, if the signal is band-limited to a maximum frequency $f_{max}$, it can be fully captured by sampling at a rate ($f_s$) such that $f_s > 2f_{max}$. The signal is sampled at discrete time instants defined as $t_n = nT$, where $T = \frac{1}{f_s}$, resulting in the sequence of samples: $x[n] = x(nT)$, $n = 0,1,2,\ldots,N-1$. Quantization then maps each continuous amplitude $x(nT)$ to a discrete value from a finite set determined by the bit-depth of the digital system (e.g., 16-bit or 32-bit resolution), which introduces a small error. The digital array of quantized samples represents the original sound in a form that can be stored as a NumPy array for further processing and analysis.

Transforming audio into an image involves encoding the waveform into pixel values while embedding essential metadata for reconstruction. Given an audio signal ($x$) with samples normalized in the range ($[-1,1]$), we map it to an 8-bit grayscale image using the transformation:

$$\text{pixel} = \left(\frac{x+1}{2}\right) \times 255$$

To ensure recoverability, a header row is embedded in the image, storing metadata such as original length ($L$) and sample rate ($SR$) as ASCII-encoded text (e.g., "L:50000;SR:22050"). The image is structured as an ($S \times S$) square, where ($S$) is chosen such that :

$$S^2 - S \geq L,$$

reserving the first row for metadata and storing audio samples in the remaining pixels. During reconstruction, the image is read, the header is parsed to extract ($L$) and ($SR$), and the audio samples are reconstructed by inverting the transformation:

$$x = \frac{\text{pixel}}{255} \times 2 - 1.$$

This approach ensures the audio signal remains intact while being robust to metadata stripping in file transfers.

*Email Address* : younessatif1.0@gmail.com

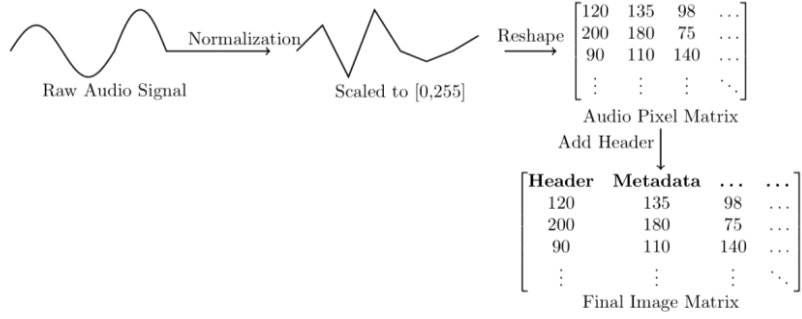

An audio sample is taken as an example, and the transformation is applied:

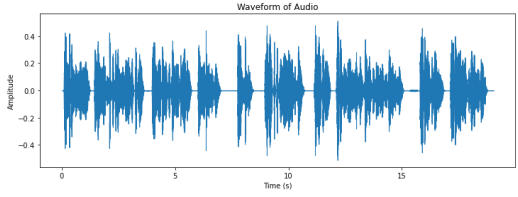

*Result after applying transformation manipulation to the audio:*

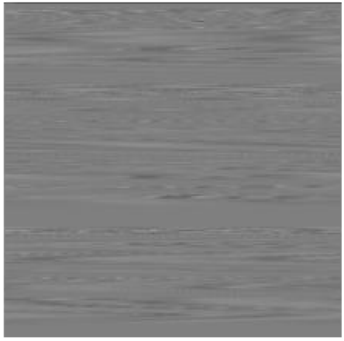

## 3   Reverse Audio Encoding

The process of reconstructing audio data from an embedded image begins with reading a grayscale image formatted as an ($S \times S$) matrix, where the first row contains metadata, including the original audio length (($L$)) and sample rate (($SR$)), stored as an ASCII-encoded header. This metadata is parsed to retrieve essential parameters for accurate audio recovery.

Following metadata extraction, the remaining rows of the image are processed to obtain the audio data. These rows contain audio samples encoded as pixel values in the range ($[0,255]$). To restore the original audio signal, the pixel values are mapped back to the normalized range ($[-1,1]$) using the transformation:

$$x = \frac{pixel}{255} \times 2 - 1$$

The reconstructed audio samples are then trimmed to the original length ($L$) to eliminate padding artifacts introduced during embedding. Finally, the extracted samples are saved in an audio format, such as WAV, using the recovered sample rate ($SR$).

This method leverages image processing techniques to encode and store audio data within digital images, ensuring signal integrity through careful normalization and metadata encoding. It presents a novel approach for audio transmission and retrieval by utilizing images as a medium for data preservation and reconstruction.

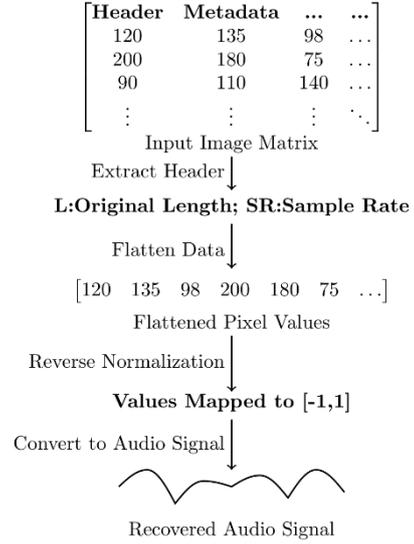

## 4   Multilayer Contributions to Fingerprint Features

### 4.1   Pitch

Pitch estimation, crucial in audio analysis, involves determining the fundamental frequency ($f_0$) of a sound, which represents its main tone. The autocorrelation method is commonly employed for this purpose, where the similarity of a signal to itself is measured across different time shifts. The first significant peak in the autocorrelation function corresponds to the period of the signal, from which the fundamental frequency is calculated by dividing the sample rate by the lag at which this peak occurs. Since pitch varies over time in real-world signals, such as speech and music, the signal is divided into small overlapping frames, and pitch is estimated for each. The mean and median of these pitch values provide important statistical measures. The mean gives an average pitch but is sensitive to noise, while the median offers a more robust estimate, less influenced by outliers. Both measures contribute valuable insights into the tonal characteristics of the signal, with the mean reflecting general pitch trends and the median focusing on the central tendency, making them essential for various applications in voice and music analysis [3].

The fundamental frequency ($f_0$) is determined by the formula:

$$f^0 = \frac{1}{\tau_{peak}} \times \text{Sample Rate}$$

where ($\tau_{peak}$) is the lag at which the first significant peak in the autocorrelation function occurs.

*Result after applying pitch manipulation to the previous audio:*

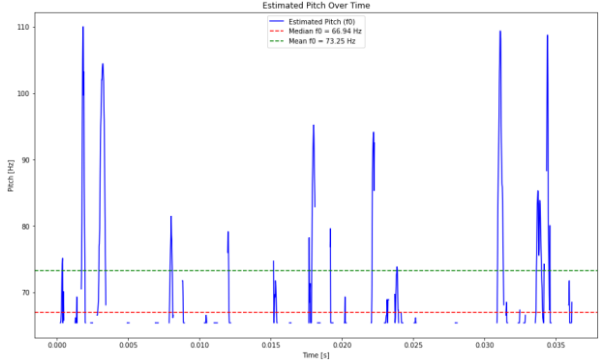

### 4.2 spectral centroid

The spectral centroid is a fundamental feature in signal processing used to quantify the brightness of a sound by representing the center of mass of the frequency spectrum. It is computed as the weighted mean frequency of the power spectrum, with the frequency magnitudes serving as weights.[4] A higher spectral centroid indicates a concentration of energy in higher frequencies, resulting in a brighter sound, while a lower centroid reflects energy in lower frequencies, creating a darker or duller sound. This feature is essential in various applications, including music information retrieval (MIR), speech processing, and emotion recognition, where it aids in distinguishing between musical genres, instruments, speakers, and emotional tones.[5] Real-time analysis of the spectral centroid provides insights into the tonal evolution of sound. The mathematical computation of the spectral centroid at a given time ($t$) is defined as:

$$C(t) = \frac{\sum_f f \cdot |X(t,f)|}{\sum_f |X(t,f)|}$$

Where ($f$) is the frequency bin index, ($X(t,f)$) is the magnitude of the frequency component at time ($t$) and frequency ($f$), and ($C(t)$) represents the centroid of the spectrum at time ($t$).

*Result after applying spectral centroid manipulation to the previous audio:*

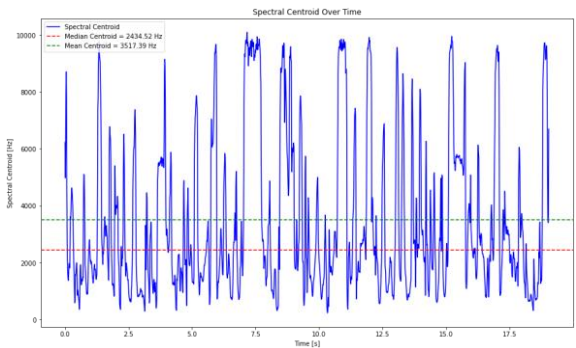

### 4.3 Spectral bandwidth

Spectral bandwidth is a key spectral feature derived from the Fourier Transform of an audio signal, representing the dispersion of energy in the frequency domain. It quantifies the "sharpness" or "dullness" of a sound by measuring the spread of frequencies around the spectral centroid, which indicates the center of mass of the frequency spectrum.[6] A high spectral bandwidth corresponds to bright, sharp sounds with a wide frequency spread, while a low spectral bandwidth is associated with more tonal, narrow-frequency sounds. Spectral bandwidth is calculated as the standard deviation of frequencies around the spectral centroid and is crucial for applications in audio classification, speech analysis, and music information retrieval (MIR) [4]. The process involves computing the Short-Time Fourier Transform (STFT) of the audio signal, calculating the spectral centroid, and then measuring the spread of frequencies relative to it. The final equation for spectral bandwidth is given by:

$$B(t) = \sqrt{\frac{\sum_f (f - C(t))^2 X(t,f)}{\sum_f X(t,f)}}$$

where ($f$) is the frequency bin index, ($C(t)$) is the spectral centroid at time ($t$), ($X(t,f)$) is the magnitude of the frequency component at time ($t$) and frequency ($f$), and ($B(t)$) is the spectral bandwidth at time ($t$).

*Result after applying Spectral Bandwidth manipulation to the previous audio:*

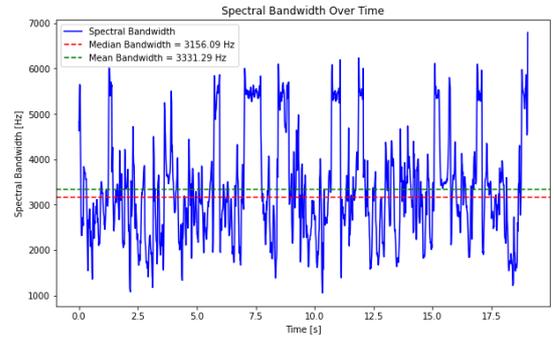

### 4.4 Spectral rolloff

Spectral rolloff is a crucial feature in audio signal processing, quantifying the frequency below which a specified percentage of the total spectral energy is contained. It serves as a distinguishing factor between harmonic and non-harmonic sounds, making it valuable for tasks such as music genre classification, speech recognition, and environmental sound analysis. Spectral rolloff is typically calculated by identifying the frequency threshold where a predetermined percentage (e.g., 85%) of the signal's energy is concentrated.[4] Harmonic sounds usually exhibit lower rolloff values, while percussive or noisy sounds tend to have higher values. In mathematical terms, spectral rolloff is defined as the frequency ($f_r$) where the cumulative sum of the Fourier magnitudes first exceeds a specific threshold, often 85% of the total spectral sum. The equation for spectral rolloff is:

$$f_r = min\left(f_k \mid \sum_{k=0}^{k_r}|X[k]| \geq \alpha \sum_{k=0}^{N-1}|X[k]|\right)$$

where ($|X[k]|$) represents the magnitude of the Fourier coefficients, ($k_r$) is the index where the cumulative energy surpasses the threshold, and ($N$) is the total number of frequency bins.

*Result after applying Spectral rolloff manipulation to the previous audio:*

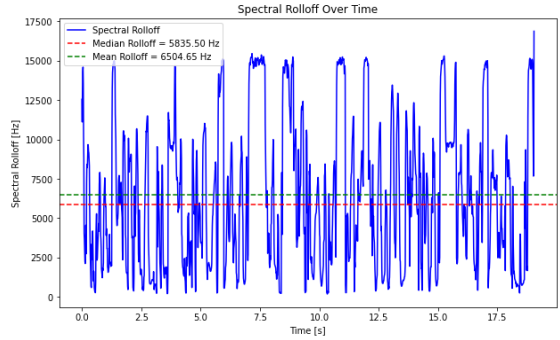

### 4.5 Zero Crossing Rate

The Zero Crossing Rate (ZCR) [3] is a key feature in audio signal processing, representing the rate at which a signal crosses the zero amplitude line. It is calculated by counting the number of sign changes in the signal within a given frame. ZCR is particularly useful in distinguishing between voiced and unvoiced sounds in speech processing, where voiced sounds exhibit lower ZCR values due to their smooth and periodic nature, while unvoiced sounds show higher ZCR values due to more erratic waveforms. Additionally, ZCR is applied in music genre classification [4], with genres like classical music showing lower ZCR and electronic or percussion-heavy genres displaying higher values. ZCR is also a useful indicator of noise, as noisy signals often have higher ZCR due to rapid fluctuations. In machine learning models, ZCR is often combined with other features, such as Mel-frequency cepstral coefficients (MFCCs), to improve audio classification tasks. Mathematically, for a discrete signal $(x[n])$, the ZCR is computed as the number of sign changes in the signal, which can be expressed as:

$$ZCR = \sum_{n=1}^{N-1} I(x[n] \cdot x[n-1] < 0)$$

where $(I)$ is the indicator function that returns 1 when the condition holds true (sign change) and 0 otherwise.

*Result after applying ZCR manipulation to the previous audio:*

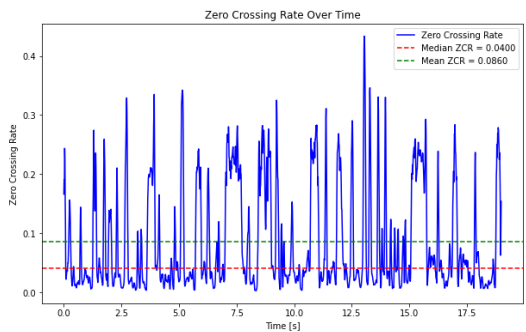

### 4.6 Mel-Frequency Cepstral Coefficients

Mel-Frequency Cepstral Coefficients (MFCCs) [7] are widely used features in speech and audio signal processing, essential for tasks like speech recognition and classification. The process of computing MFCCs involves several key steps: first, the audio signal is transformed into the frequency domain using the Discrete Fourier Transform (DFT), followed by the application of the Mel scale to mimic human auditory perception. The signal is then passed through a Mel filterbank, which filters the power spectrum into Mel frequency bands. A logarithmic transformation of the energy in each Mel band simulates the ear's response to loudness, and the Discrete Cosine Transform (DCT) is applied to decorrelate the energies, resulting in uncorrelated cepstral coefficients. The final MFCCs, typically the first 13 coefficients, capture the most significant spectral features of the signal. The process can be summarized as follows:

$$C_n = \sum_{m=0}^{M-1} log(E_m) \cdot cos\left(\frac{\pi n}{M}\left(m + \frac{1}{2}\right)\right)$$

where $(C_n)$ are the MFCCs, $(E_m)$ represents the Mel-filtered energies, and $(M)$ is the number of Mel bands.

*Result after applying MFCCs manipulation to the previous audio:*

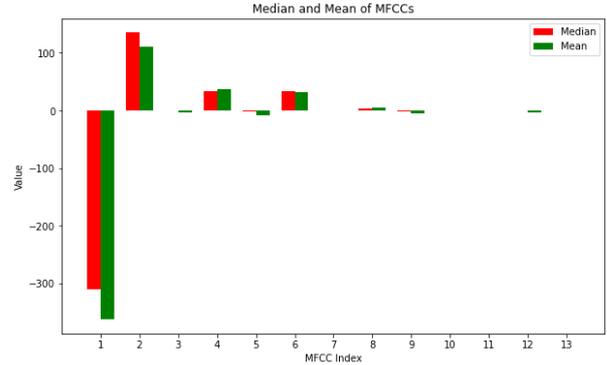

### 4.7 RMS energy

RMS energy (loudness) is a widely employed method in audio signal processing for measuring the loudness or power of an audio signal.[8] It is calculated by computing the Root Mean Square (RMS) of the signal's amplitude over a specific time frame. The process involves squaring the signal values, averaging them, and taking the square root of that average, providing a time-varying measure of loudness. Higher RMS values correspond to louder sections, while lower values represent quieter ones. RMS energy is commonly used in fields such as speech and music analysis, where it helps detect variations in loudness and assess the dynamic range. However, it does not account for frequency-dependent loudness or human auditory perception. Despite these limitations, it remains a fundamental tool for analyzing audio signals. Mathematically, for a discrete signal $(x[n])$, the RMS energy is expressed as:

$$\text{RMS Energy}(x) = \sqrt{\frac{1}{N}\sum_{n=0}^{N-1} x[n]^2}$$

*Result after applying RMS energy manipulation to the previous audio:*

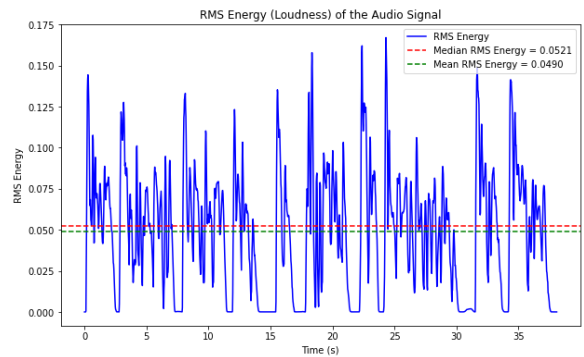

### 4.8 Harmonic-to-Noise Ratio

Harmonic-to-Noise Ratio (HNR) [9] is a spectral feature that quantifies the balance between harmonic (periodic) and noise (aperiodic) components in an audio signal, providing an indication of sound quality. A higher HNR corresponds to a cleaner, more periodic signal, typical of voiced speech or musical tones, while a lower HNR suggests a noisier, less periodic signal, often seen in unvoiced speech or background noise. HNR is calculated by comparing the Root Mean Square (RMS) energy of the harmonic and noise components, and it is widely used in speech processing, music analysis, and clinical settings to assess voice disorders.[10] The HNR can be computed as the ratio of RMS energy of the harmonic component to the RMS energy of the noise component:

$$HNR = \frac{RMS_{harmonic}}{RMS_{percussive}}$$

*Result after applying HNR manipulation to the previous audio:*

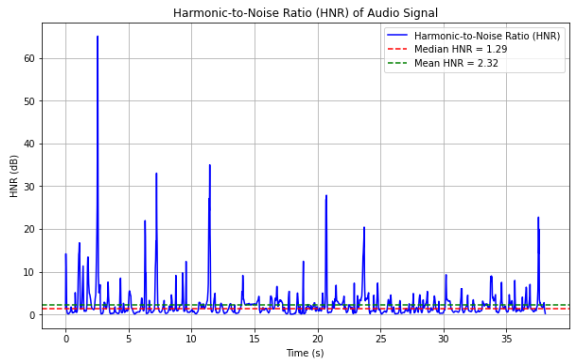

### 4.9 Spectral flatness

Spectral flatness is a key measure in audio signal processing used to assess how noise-like or tonal a sound is.[8] It quantifies the distribution of energy across different frequencies by comparing the geometric and arithmetic means of the power spectrum of a signal. A spectral flatness value close to zero indicates a tonal signal, where energy is concentrated in specific frequencies, while a value near one suggests a noise-like signal, where energy is more uniformly distributed. This measure is commonly applied in speech and music analysis, audio classification, and other tasks such as music genre recognition and environmental sound classification. It is computed by first applying the Short-Time Fourier Transform (**STFT**) [11] to the signal to obtain its power spectrum, followed by calculating the geometric and arithmetic means for each frequency bin. The spectral flatness **SF** is given by the following equation:

$$SF = \frac{(\prod_{k=0}^{N-1} X[k])^{\frac{1}{N}}}{\frac{1}{N}\sum_{k=0}^{N-1} X[k]}$$

where $(X[k])$ represents the power spectrum at frequency bin $(k)$, and $(N)$ is the total number of frequency bins.

*Result after applying Spectral flatness manipulation to the previous audio:*

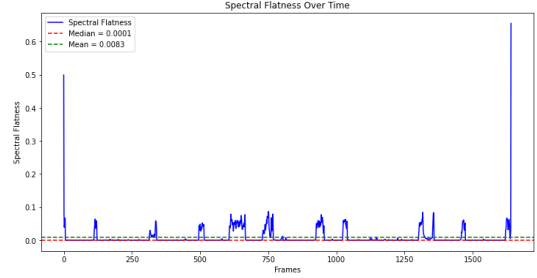

### 4.10 Spectral contrast

Spectral contrast [4][12] is a crucial feature in audio signal processing that quantifies the variation between the peaks and valleys in the power spectrum across different frequency bands. Unlike other spectral features that describe the overall energy distribution, spectral contrast focuses on the variation within specific frequency ranges by dividing the spectrum into subbands and calculating the logarithmic ratio of the highest and lowest energy levels within each band. This feature is effective in distinguishing harmonic from non-harmonic components of a signal and is widely used in applications such as speech recognition, music analysis, and audio classification. It is particularly useful for characterizing timbres of musical instruments, identifying noisy or reverberant environments, and differentiating percussive sounds from speech. The spectral contrast for a given frequency band ($b$) is mathematically defined as:

$$SC(b) = 10 \cdot log_{10}\left(\frac{P_{max}(b)}{P_{min}(b)}\right)$$

where $(P_{max}(b))$ and $(P_{min}(b))$ represent the maximum and minimum energy levels within the band, respectively.

*Result after applying Spectral contrast manipulation to the previous audio:*

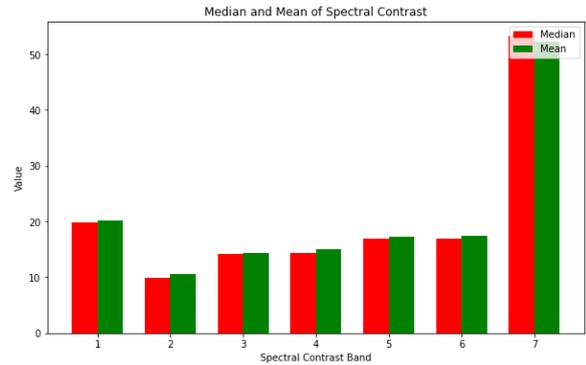

### 4.11 Chroma

Chroma features are essential in audio analysis, particularly for tasks such as music key detection, chord recognition, and genre classification.[4][13] These features capture the harmonic and tonal content of a signal by representing the distribution of spectral energy across the 12 pitch classes (C, C#, D, D#, E, F, F#, G, G#, A, A#, B), while ignoring octave information. The chroma representation is derived from the short-time Fourier transform (STFT) of the signal, with the energy of spectral components corresponding to the same pitch class summed across octaves. This approach provides a compact yet informative representation of harmonic content, which is perceptually significant since notes with the same chroma share tonal similarities. Chroma features are commonly used in applications like music retrieval and speech emotion recognition, with their visualization through chromagrams aiding the interpretation of harmonic patterns over time. To compute chroma features, the following equation is used:

$$C(n) = \sum_{k \in \mathcal{K}(n)} |X[k]|$$

where $C(n)$ represents the chroma energy for pitch class $n$, $X[k]$ is the magnitude spectrum at frequency bin $k$, and $\mathcal{K}(n)$ is the set of frequency bins corresponding to pitch class $n$ across octaves.

Result after applying Chroma manipulation to the previous audio:

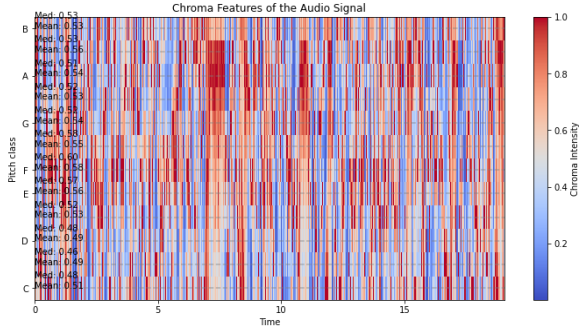

## 5 Fingerprint Embedded into Replicated Structure in the Red Channel

The proposed method extracts a set of acoustic features from an audio signal, including both scalar features (such as median and mean fundamental frequency, spectral centroid, spectral bandwidth, spectral rolloff, and zero-crossing rate) and vector features (such as MFCCs, spectral contrast, and chroma). These features, along with additional measures like RMS energy and harmonic-to-noise ratio, are combined into a one-dimensional array. To address the varying scales of the features, min-max normalization is applied to each element, ensuring the values are mapped to the range [0, 255]. The normalized array is then replicated to achieve a target length of 262144 elements, after which it is reshaped into a 512×512 matrix to form the red channel of an RGB image. This transformation embeds the acoustic fingerprint into a fixed spatial structure, which can be used for further analysis.

The normalization process is mathematically represented as:

$$x_i' = \frac{x_i - \min(\mathbf{x})}{\max(\mathbf{x}) - \min(\mathbf{x}) + \varepsilon} \times 255,$$

where $\varepsilon$ is a small constant to avoid division by zero.

*Expected image:*

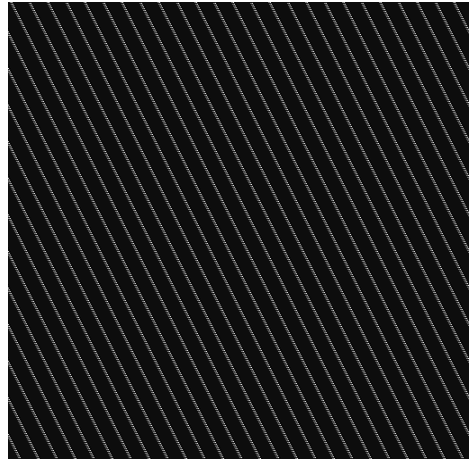

This method effectively bridges heterogeneous acoustic features into a unified, fixed-dimensional representation, which is then used to form a spatial "fingerprint" image. Such an image is well-suited for analysis by image-based machine learning models, including convolutional neural networks, thereby facilitating tasks like voice recognition and classification.

*Result after applying Algorithm 1 to the previous audio:*

## 6 Fingerprint Embedded into Square Patches in the Blue Channel

The proposed method for constructing a fingerprint layer involves first transforming heterogeneous acoustic features into fixed-size, spatially localized patches and then assembling these patches into a composite image. For each scalar feature—such as the median and mean of the fundamental frequency, spectral centroid, spectral bandwidth, spectral rolloff, zero-crossing rate, RMS energy, and harmonic-to-noise ratio—a linear transformation is applied to normalize the value to the [0,255] range. Each scalar is then represented as a patch, where, for example, the left half of a $Z \times Z$ block is filled with the normalized median and the right half with the normalized mean. For vector features such as MFCCs, spectral contrast, and chroma, the individual components are first normalized based on their minimum and maximum values; these vectors are then linearly interpolated to generate rows of length $Z$, and tiled vertically to form a $Z \times Z$ patch. Mathematically, if a scalar $f$ is normalized to $f_{\text{norm}} \in [0,255]$, the corresponding patch $P$ is defined as

$$P(i,j) = f_{\text{norm}}, \quad \forall\, 1 \leq i, j \leq Z.$$

Similarly, for a vector $\boldsymbol{v}$ of length $L$, the normalized vector is interpolated over a new set of $Z$ points, and then tiled to form a patch $P \in R^{Z \times Z}$. Once all patches are generated, they are arranged into a grid that fills a 512 × 512 image. If $N$ patches are obtained, grid dimensions $r$ and $c$ are chosen such that $r \times c \geq N$ and the patch size is determined by :

$$(Z = \frac{512}{c})$$

Any unfilled grid cells can be assigned a default value, such as 127 (mid-gray). This approach effectively embeds the voice's multidimensional fingerprint into a fixed spatial layout that can be directly fed into convolutional neural networks for voice recognition and classification tasks.

*Expected image:*

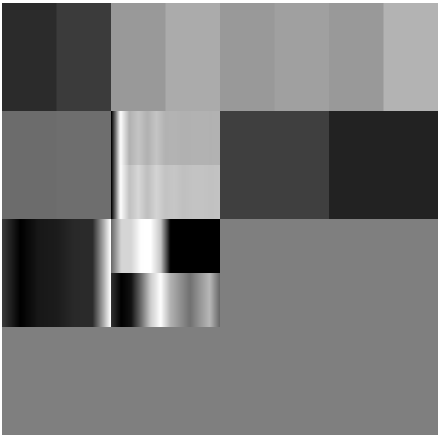

This approach enables each feature to contribute a visually distinct patch within the composite image, thereby embedding the voice's multidimensional fingerprint into a format that can be directly processed by convolutional neural networks for tasks such as voice recognition and classification.

*Result after applying Algorithm 2 to the previous audio:*

patches, where the left and right halves contain the normalized median and mean, respectively:

$$P(i,j) = \begin{cases} f_{norm, median}, & if\ j \leq Z/2, \\ f_{norm, mean}, & if\ j > Z/2, \end{cases} \forall 1 \leq i,j \leq Z.$$

These patches are arranged in a grid to form the blue channel $I_{\text{blue}}$. The final fingerprint image is generated by merging all three channels:

$$I_{\text{RGB}} = \text{merge}(I_{\text{red}}, I_{\text{green}}, I_{\text{blue}}).$$

This structured encoding preserves both temporal and spectral information, facilitating effective processing by convolutional neural networks for voice recognition and classification tasks.

*Expected image:*

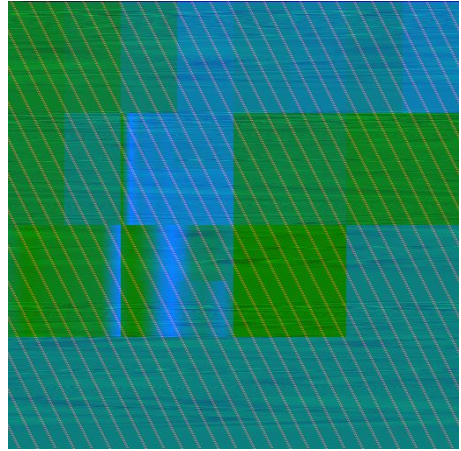

*Result after applying Algorithm 3 to the previous audio:*

## 7 Multi-Channel Audio Fingerprint Representation for Voice Recognition

The proposed method constructs a multi-channel RGB image that encapsulates both the raw audio waveform and its acoustic features, creating a comprehensive fingerprint for analysis. The grayscale channel ($I_{\text{green}} \in R^{512 \times 512}$) represents the temporal dynamics of the waveform, while the red and blue channels encode high-level audio features. Specifically, extracted features such as the fundamental frequency, spectral attributes, MFCCs, and harmonic-to-noise ratio are flattened, normalized to [0,255], and tiled into $I_{\text{red}}$, ensuring a structured representation. Simultaneously, these features are transformed into $Z \times Z$

## 8 Proposed Model for Audio-Based Fingerprint Representation

In this work, a deep convolutional neural network designed to recognize speakers using image representations of audio fingerprints. The model processes 512×512 RGB images, where each image encodes rich acoustic features extracted from the audio signal. To enhance robustness and mitigate overfitting, a data augmentation pipeline is applied that randomly flips images horizontally, rotates them by 0.1 radians, and performs random zooming by 10%. The network architecture begins with an input layer that explicitly defines the input shape, followed by the data augmentation block. This is succeeded by four convolutional blocks where the first block employs 32 filters of size 3×3, and subsequent blocks increase the filter count to 64, 128, and 256 respectively; each convolutional layer is immediately followed by batch normalization and max pooling to stabilize training and reduce spatial dimensions. The learned feature maps are then flattened and passed through a dense layer with 256 neurons using ReLU activation. A dropout layer with a 50% rate is employed to further prevent overfitting before the final softmax classification layer, which outputs predictions across the target speaker classes. The model is compiled with the Adam optimizer (learning rate = 0.0001) and trained using categorical crossentropy loss. The training dataset consists of 1096 audio (images) (derived from a total of 548 audio samples per speaker), while the remaining 48 samples per speaker serve as the evaluation set. This architecture not only leverages convolutional layers to capture spatial hierarchies in the audio-derived images but also integrates data augmentation and regularization strategies to ensure recognition tasks.

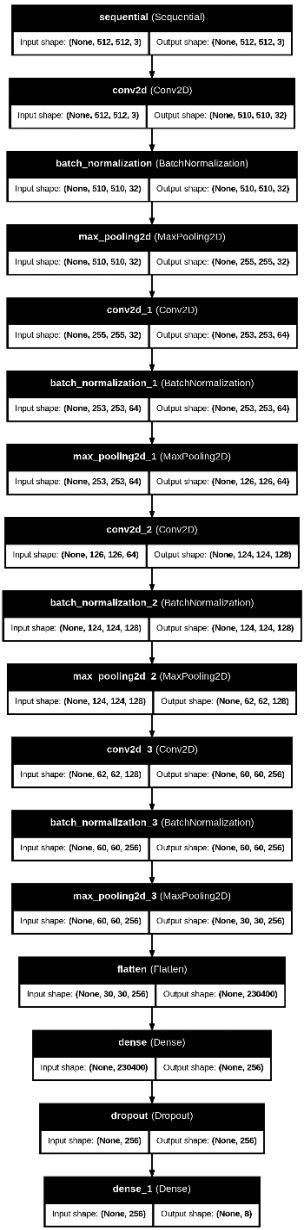

**Figure 1**: Architecture of the Proposed Audio-Based Fingerprint Representation Model

## 9    Training and Validation Accuracy Analysis

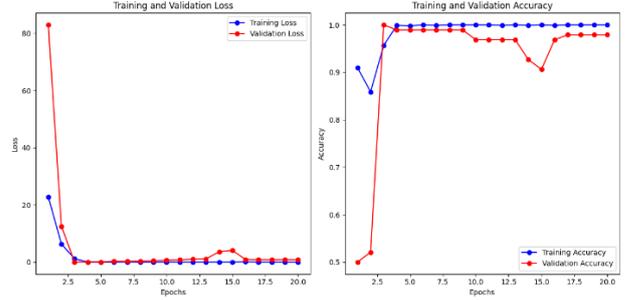

**Figure 2**: Training and Validation Accuracy

The training and validation accuracy curves across 20 epochs reveal insightful trends regarding model performance. Initially, the model exhibits significant fluctuations, with the training accuracy starting at 92.34% and the validation accuracy at only 50.00% during the first epoch. Notably, by the third epoch, the training accuracy rapidly improves to 85.86% while the validation accuracy unexpectedly reaches nearly 100%, suggesting an early convergence on certain features of the data. As training progresses, the model consistently achieves near-perfect accuracy on the training set, with loss values approaching zero. However, the validation accuracy, while remaining high (approximately 97.92% by the final epoch), shows minor oscillations in loss values, particularly after the tenth epoch, indicating subtle variations in generalization performance. These patterns, illustrated in the corresponding diagrams, underscore that the model is effectively learning the complex features inherent in the audio-based fingerprint representations, albeit with occasional fluctuations that may warrant further investigation into regularization or hyperparameter tuning.

## 10    Experimental Results and Performance Analysis

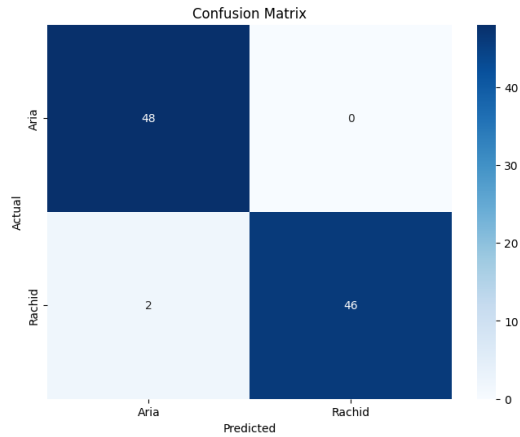

**Figure 3**: Confusion Matrix

The proposed model demonstrates exceptional performance in speaker recognition, achieving a test accuracy of 97.92% on the evaluation set. The overall classification metrics further reinforce its efficacy, with both speakers attaining precision, recall, and F1-scores near 0.98. These results indicate that the model effectively distinguishes between the two speaker classes, highlighting the robustness of the image-based audio fingerprint representation.

The high accuracy and balanced classification metrics underscore the potential of this approach in practical voice recognition applications, providing a promising direction for further research and deployment in real-world scenarios.

|  | precision | recall | f1-score | support |
|---|---|---|---|---|
| **Aria** | 0.96 | 1.00 | 0.98 | 48 |
| **Rachid** | 1.00 | 0.96 | 0.98 | 48 |
|  |  |  |  |  |
| **accuracy** |  |  | 0.98 | 96 |
| **macro avg** | 0.98 | 0.98 | 0.98 | 96 |
| **weighted avg** | 0.98 | 0.98 | 0.98 | 96 |

**Table 1**: Classification Report

## 11 Discussion and Interpretation of Findings

The current training results, while promising, underscore the limitations inherent in utilizing a relatively small dataset comprising only two speakers. Expanding the dataset to include a more diverse and extensive pool of speakers is expected to significantly enhance the model's generalization capabilities and robustness in real-world scenarios. Future work should focus on collecting and curating a comprehensive audio dataset that spans various demographics, accents, and recording conditions, which could provide the necessary variability to fine-tune and further optimize the model. In addition, exploring advanced data augmentation techniques and leveraging transfer learning from larger, pre-trained models may further improve performance and mitigate the challenges associated with limited training data.

## 12 Conclusion and Future Perspectives

This study has demonstrated a novel approach to voice recognition by converting audio signals into image-based fingerprint representations and processing them using a deep convolutional neural network. Despite the promising accuracy achieved on a limited dataset of two speakers, the results highlight the potential benefits of expanding the training corpus to encompass a broader and more diverse range of voices. Future work should focus on curating larger, more varied datasets and integrating advanced augmentation and transfer learning techniques to further improve generalization and robustness. Overall, the findings pave the way for innovative applications in audio analysis, with the prospect of enhancing speaker recognition systems in real-world scenarios through continued methodological refinement and dataset expansion.

**Data Availability Statement**

ALGORITHMS:

**Algorithm 1**: Feature Replication and Embedding Algorithm (FREA)

*START*
1. features = extract_voice_features(audio)
2. vector_list = []
3. *FOR* each feature *IN* features *DO*
4.     *IF* feature is scalar *THEN*
5.         append feature to vector_list
6.     *ELSE IF* feature is a vector *THEN*
7.         *FOR* each element *IN* feature *DO*
8.             append element to vector_list
9.         *ENDFOR*
10.     *ENDIF*
11. *ENDFOR*
12. flat_array = convert vector_list to 1D array
13. min_val = MIN(flat_array), max_val = MAX(flat_array)
14. normalized_array = (flat_array - min_val) / (max_val - min_val + ε) * 255
15. target_size = 512 * 512
16. replicated_array = tile(normalized_array) until LENGTH(replicated_array) ≥ target_size
17. truncated_array = first target_size elements of replicated_array
18. image_matrix = reshape(truncated_array, (512, 512))

*END*

**Algorithm 2**: Voice Fingerprint in Blue Channel Square Patches Embedding Algorithm

*START*
1. features = extract_voice_features(audio_file)
2. patches = []
3. *FOR* each (key, value) *IN* features *DO*
4.     *IF* value is a scalar pair (e.g., median and mean) *THEN*
5.         norm_median = normalize_scalar(median_value, min_val, max_val)
6.         norm_mean = normalize_scalar(mean_value, min_val, max_val)
7.         patch = create_scalar_patch(norm_median, norm_mean, patch_size)
8.     *ELSE IF* value is a vector pair *THEN*
9.         norm_vec1 = normalize_vector(median_vector)
10.         norm_vec2 = normalize_vector(mean_vector)
11.         patch = create_vector_patch(norm_vec1, norm_vec2, patch_size)
12.     *END IF*
13. *ENDFOR*
14. grid_rows = number of rows (e.g., 4)
15. grid_cols = number of columns (e.g., 4)
16. patch_size = Z   // e.g., 128 pixels so that grid_rows*Z = 512
17. final_image = matrix of size (grid_rows * patch_size, grid_cols * patch_size) with default value
18. index = 0
19. *FOR* row *from* 0 *to* grid_rows - 1 *DO*
20.     *FOR* col *from* 0 *to* grid_cols - 1 *DO*
21.         *IF* index < length(patches) *THEN*
22.             Insert patches[index] into final_image at:
                  [row * patch_size : (row+1) * patch_size, col * patch_size : (col+1) * patch_size]
23.             index = index + 1
24.         *ELSE*
25.             // Optionally leave remaining cells filled with default value
26.         *END IF*
27.     *ENDFOR*
28. *ENDFOR*
29. *RETURN* final_image

*END*

| **Algorithm 3**: Voice Fingerprint RGB Fusion Algorithm |
|---|
| *START* |
| **1**      I_green = GenerateGrayscaleImageFromAudio(audio) |
| **2**      features = extract_voice_features(audio) |
| **3**      flat_features = Flatten(features) |
| **4**      normalized_features = Normalize(flat_features, [0, 255]) |
| **5**      I_red = ReplicateAndReshape(normalized_features, (512, 512)) |
| **6**      patches = [] |
| **7**      *FOR* each feature pair (median, mean) *IN* features *DO* |
| **8**          patch = CreatePatch(feature.median, feature.mean, patch_size) |
| **9**          *Append* patch to patches |
| **10**      *ENDFOR* |
| **11**      I_blue = AssemblePatchesIntoGrid(patches, (512, 512)) |
| **12**      I_RGB = MergeChannels(I_red, I_green, I_blue) |
| **13**      *RETURN* I_RGB |
| *END* |

# 13 References:

[1] Sur, M. (2000). MIT researcher finds that part of brain used for hearing can learn to 'see'. *Nature*, April 20, 2000.

[2] Atif, Y. (2025, March 5). Audio Classification Model. Kaggle. http://www.kaggle.com/code/atif10/audio-classification-model

[3] Rabiner, L. R., & Schafer, R. W. (1978). *Digital Processing of Speech Signals*. Englewood Cliffs, NJ: Prentice-Hall.

[4] Tzanetakis, G., & Cook, P. (2002). *Musical Genre Classification of Audio Signals*. IEEE Transactions on Speech and Audio Processing, 10(5), 293–302.

[5] Casey, M., Veltkamp, R., Goto, M., Leman, M., Rhodes, C., & Sloboda, J. (2008). *Content-based music information retrieval: Current directions and future challenges*. Proceedings of the IEEE, 96(4), 668–696.

[6] Peeters, G. (2004). "A large set of audio features for sound description (similarity and classification) in the CUIDADO project." IRCAM Technical Report.

[7] Davis, S., & Mermelstein, P. (1980). *Comparison of parametric representations for monosyllabic word recognition in continuously spoken sentences*. IEEE Transactions on Acoustics, Speech, and Signal Processing, 28(4), 357–366.

[8] Oppenheim, A. V., & Schafer, R. W. (2010). *Discrete-Time Signal Processing* (3rd ed.). Prentice Hall.

[9] Boersma, P., & Weenink, D. (2001). *Praat: doing phonetics by computer* [Computer software]. Retrieved from http://www.praat.org/

[10] Hillenbrand, J., Cleveland, R. A., & Erickson, R. L. (1994). Acoustic correlates of breathy vocal quality. *Journal of Speech, Language, and Hearing Research, 37*(4), 769–778.

[11] Smith, J. O. (2011). *Spectral Audio Signal Processing*. W3K Publishing. Available online: http://www.jos.ito/freebooks/sasp/

[12] Müller, M. (2015). *Fundamentals of Music Processing*. Springer.

[13] Fujishima, T. (1999). *Realtime Chord Recognition of Musical Audio*. In Proceedings of the International Conference on Music Information Retrieval (ISMIR).

- Youness Atif, "Voices as Images". Zenodo, March 07, 2025. doi: 10.5281/zenodo.14988955.

- Youness Atif, "Audio Classification Model". Zenodo, March 07, 2025. doi: 10.5281/zenodo.14989013.